\documentclass[aps,prl,preprint,superscriptaddress]{revtex4-1}

\usepackage{subfigure}
\usepackage{epsfig}
\usepackage{float}
\usepackage{multirow}
\usepackage{url}

\usepackage{graphicx}
\usepackage[colorlinks,urlcolor=blue,linkcolor=blue,anchorcolor=blue,citecolor=blue]{hyperref}
\usepackage{amssymb, amsmath, amsthm}
\usepackage[usenames]{color}
\usepackage{dcolumn}
\usepackage{bm}
\usepackage{ textcomp }
\usepackage[normalem]{ulem}

\begin{document}
\title{Effective Model for Fractional Topological Corner Modes in Quasicrystals}

\author{Citian Wang}
\affiliation{School of Physics, Peking University, Beijing 100871, China}

\author{Feng Liu}
\email[Corresponding author: ]
{fliu@eng.utah.edu}
\affiliation{Department of Materials Science and Engineering, University of Utah, Salt Lake City, Utah 84112, USA}

\author{Huaqing Huang}
\email[Corresponding author: ]
{huaqing.huang@pku.edu.cn}
\affiliation{School of Physics, Peking University, Beijing 100871, China}
\affiliation{Collaborative Innovation Center of Quantum Matter, Beijing 100871, China}
\affiliation{Center for High Energy Physics, Peking University, Beijing 100871, China}

\date{\today}

\begin{abstract}
High-order topological insulators (HOTIs), as generalized from topological crystalline insulators (TCIs), are characterized with lower-dimensional metallic boundary states protected by spatial symmetries of a crystal, whose theoretical framework based on band inversion at special $k$-points cannot be readily extended to quasicrystals because quasicrystals contain rotational symmetries that are not compatible with crystals, and momentum is no longer a good quantum number. Here, we develop a low-energy effective model underlying HOTI states in 2D quasicrystals for all possible rotational symmetries. By implementing a novel Fourier transform developed recently for quasicrystals and approximating the long-wavelength behavior by their large-scale average, we construct an effective $k \cdot p$ Hamiltonian to capture the band inversion at the center of a pseudo-Brillouin zone (PBZ). We show that an in-plane Zeeman field can induce mass-kinks at the intersection of adjacent edges of a 2D quasicrystal TI and generate corner modes (CMs) with fractional charge, protected by rotational symmetries. Our model predictions are confirmed by numerical tight-binding calculations. Furthermore, when the quasicrystal is proximitized by an \textit{s}-wave superconductor, Majorana CMs can also be created by tuning the field strength and chemical potential. Our work affords a generic approach to studying the low-energy physics of quasicrystals, in association with topological excitations and fractional statistics.
\end{abstract}


\maketitle

\textit{Introduction.}---
A first-order $\mathbb{Z}_2$ TI is characterized with a correspondence between a $d$-dimensional gapped bulk state and a $(d-1)$-dimensional gapless boundary state protected by time-reversal (TR) symmetry, while a HOTI is instead featured with a $(d-2)$-dimensional gapless boundary state protected by spatial symmetries of a crystal, such as mirror and rotation symmetry~\cite{benalcazar2017quantized,schindler2018higher,PhysRevLett.119.246401,PhysRevLett.119.246402,
schindler2018bismuth,yue2019symmetry,PhysRevLett.122.256402,PhysRevLett.124.136407,
PhysRevLett.123.256402,lee2020two,liu2019two,PhysRevLett.123.216803,PhysRevLett.126.066401,
huang2021structrual,huanghq2022pan,
PhysRevLett.120.026801,PhysRevLett.122.086804, PhysRevLett.123.186401,hsu2019topology, PhysRevLett.124.166804,PhysRevLett.125.056402,zhao2021higher}.
In the context of crystalline symmetry protected topological boundary states, the HOTI can be viewed as a generalization of TCI \cite{PhysRevLett.106.106802}. In general, $\mathbb{Z}_2$ invariant can be calculated by the product of band inversion indices at the TR invariant momenta (TRIM) in the subspace of occupied bands, the eigenvalues of inversion symmetry operator, in the first BZ \cite{parity}. One can generalize this scheme to identify a HOTI by calculating the topological invariant, the eigenvalues of a spatial symmetry at all $\mathbf{k}$-points linked by this spatial symmetry \cite{benalcazar2017quantized,schindler2018higher}. Namely, a high-order topology is defined in the subspace of a crystal operated by spatial symmetries. Apparently, this approach is not applicable to quasicrystals which do not have a BZ and momentum ($\mathbf{k}$) is no longer a good quantum number. Moreover, quasicrystals contain rotation symmetries not compatible with translational symmetry.

Alternatively, a HOTI can be viewed by gapping the $(d-1)$-dimensional gapless boundary (surfaces or edges) of a $\mathbb{Z}_2$  TI, but the band degeneracy is locally protected at the $(d-2)$-dimensional boundary (hinges and corners) by spatial symmetry. For example, in a 2D HOTI, corner modes (CMs) can be viewed as topological Jackiw-Rebbi domain-wall states \cite{jackiw1976solitons}, with opposite Dirac masses between two edges enforced by a mirror symmetry \cite{PhysRevB.97.205135,PhysRevX.9.011012,trifunovic2020higher}, and a variety of 2D HOTI systems have been proposed by implementing such mirror-invoked mass-inversion mechanism \cite{PhysRevLett.120.026801,PhysRevLett.122.086804, PhysRevLett.123.186401,hsu2019topology, PhysRevLett.124.166804, PhysRevLett.125.056402, zhao2021higher}, including interestingly CMs in quasicrystals \cite{PhysRevLett.123.196401, PhysRevLett.124.036803, PhysRevB.102.241102,huang2021generic,huang2022topological}. More generally, a CM of domain-wall state can be protected by rotation symmetry, as derived from edge network theory \cite{PhysRevB.99.155102}. Instead of opposite edge masses encoded by mirror eigenvalues ($\pm 1$), an edge-dependent Dirac masses emerges with a phase difference of $2\pi/n$, termed as mass kink \cite{PhysRevLett.47.986,PhysRevLett.50.439}, defined by eigenvalues of the $C_n$ rotation $(e^{\frac{2 m \pi i}{n}},\, m=1,\dots,n)$, giving rise to a fractional charged CMs of $e/n$. We note that the mass inversion mechanism, as also applied to quasicrystals \cite{PhysRevLett.123.196401,PhysRevLett.124.036803, PhysRevB.102.241102,huang2021generic,huang2022topological}, is a special case of mass link with $C_2$ rotation with a phase shift of $\pi$. Therefore, it is very interesting to explore if the mass kink approach can be generalized to quasicrystals, because they have rotation symmetries (such as 5-fold rotation) that are not compatible with translational symmetry, which may hinder the realization of topological CMs with fractional charges that do not exist in crystals.

\begin{figure}
\includegraphics[width =1\columnwidth]{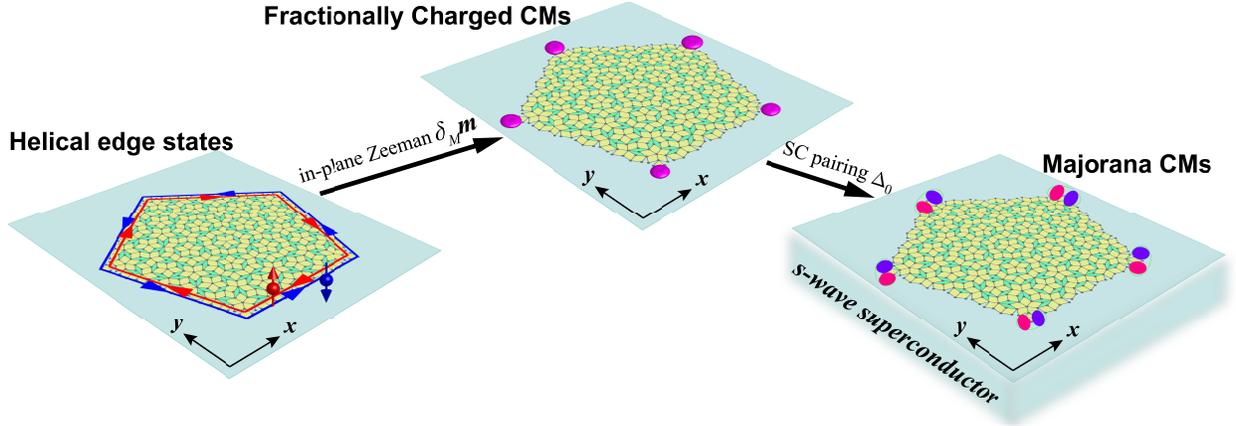}
\caption{\label{fig1} Schematic illustration of the Zeeman-field-induced topological phase transitions in a 2D pentagonal quasicrystal. Starting from a TI phase with helical edge states, the quasicrystal is driven to a HOTI phase with five charged corner modes (CMs) by an in-plane Zeeman field $\delta_M\bm{m}$. When the system is in proximity with an \textit{s}-wave superconductor (SC) with pairing $\Delta_0$, Majorana CMs can be generated by tuning the Zeeman field and chemical potential.}
\end{figure}

In this Letter, we develop a low-energy effective model for quasicrystals, based on a novel Fourier transform developed recently \cite{JIANG2014428} by representing a quasicrystal as a projection of a hypercrystal from higher dimensions, from a 4D hyper-BZ to a 2D PBZ. Then, an effective $k \cdot p$ Hamiltonian is constructed at the center ($\Gamma$) of the PBZ, under long-wavelength approximation by large-scale average with quasicrystalline symmetry. As band inversion occurs at $\Gamma$, charged and Majarona CMs arising from {fractional mass kinks} can emerge by applying an in-plane Zeeman field. Taking the Penrose-tiling quasicrystal TI as an example, we show that the field induces a fractional mass kink with a phase shift of $2\pi/5$ at the intersection of adjacent edges, generating five in-gap localized CMs with a fractional charge of $e/5$, as displayed in Fig.~\ref{fig1}. Our scheme can be easily generalized to $q$-fold quasicrystals ($q=5,8,10,12,\cdots$), leading to the corner charge fractionalization by $e/q$ that is disallowed in crystals. In addition, when the 2D quasicrystal TI is in proximity with an \textit{s}-wave SC, Majorana CMs can be generated by tuning the in-plane Zeeman field and chemical potential (see Fig.~\ref{fig1}).

\textit{Model.}---
We construct quasicrystal lattices based on the rhombic Penrose and Ammann-Beenker tilings which have 5- and 8-fold rotational symmetry, respectively \cite{penrose1974role,beenker1982algebraic}. We assume the atoms have three atomic orbitals ($s,p_x, p_y$) at the vertices of tiling and consider hoppings between neighboring vertices connected by edges or the shorter diagonals of the rhombi.
The {tight-binding (TB)} Hamiltonian is given by
\begin{eqnarray}
\label{eq1}
H&=&\sum_{i\alpha}\epsilon_\alpha c_{i\alpha}^\dag c_{i\alpha}+\sum_{\langle i\alpha,j\beta\rangle}t_{\alpha,\beta}(\mathbf{r}_{ij})c_{i\alpha}^\dag c_{j\beta}\\
 & &+i\lambda \sum_{i} (c_{ip_y}^\dag s_z c_{ip_x}-c_{ip_x}^\dag s_z c_{ip_y})\nonumber+\sum_{i\alpha} \delta_\alpha c_{i\alpha}^\dag (\bm{m}\cdot\bm{s}) c_{i\alpha}, \nonumber
\end{eqnarray}
where $c_{i\alpha}^\dag=(c_{i\alpha\uparrow}^\dag,c_{i\alpha\downarrow}^\dag)$ is electron creation operators on the $\alpha (=s,p_x,p_y)$ orbital at the $i$-th site, $\epsilon_\alpha$ is the on-site energy, and $t_{\alpha,\beta}(\mathbf{r}_{ij})$ is the Slater-Koster hopping integral which depends on the orbital type ($\alpha$ and $\beta$) and the vector $\mathbf{r}_{ij}$ between sites $i$ and $j$. $\lambda$ is the spin-orbit coupling (SOC) strength and $\bm{s}=(s_x,s_y,s_z)$ are the Pauli matrices. The last term represents a Zeeman field along the direction of $\bm{m}$. $\delta_\alpha$ depends on field strength and the g-factor of $\alpha$ orbital. For simplicity, we take a uniform value $\delta_\alpha=\delta_M$, which would not change the main physics we discuss here. Experimentally, the Zeeman term can be introduced by a magnetic field, coupling to a magnetic substrate, or depositing magnetic adatoms on quasicrystal substrates. It is well known that by considering a band inversion between \textit{s} and ($p_x$, $p_y$) orbitals, topological states can be realized in quasicrystal lattices \cite{huanghqQLprl,huanghqQLprb,huanghq2019Comparison}. Hence, we will use the same settings and focus on $n_{occ}=2/3$ filling of electron states hereafter.

\begin{figure}
\includegraphics[width =1\columnwidth]{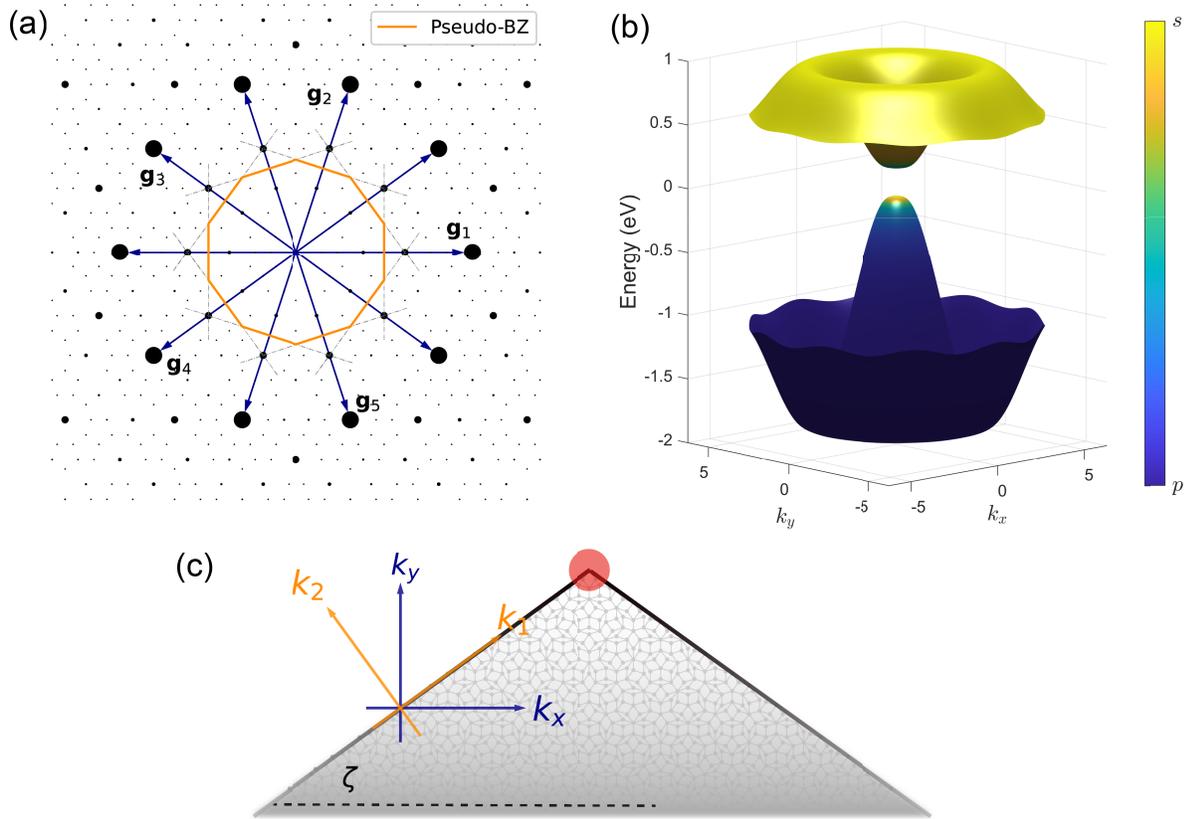}%
\caption{\label{fig2} (a) The set of all reciprocal wave vectors consists of five wave vectors $\mathbf{g}_j$, forming a dense set of points in k space. The perpendicular bisectors of the 10 principal wave vectors $\pm \mathbf{g}_j$ form a decagonal boundary to the pseudo-Brillouin zone (PBZ). (b) The effective band structure around the center of the PBZ of the Penrose-tiling quasicrystal.
(c) Schematic illustration of an edge along {$\vec{e}_{1}=\cos \zeta \hat{e}_{x}+\sin \zeta \hat{e}_{y}$} direction, where $\zeta$ is the angle between the edge and positive \textit{x} direction.} 
\end{figure}

\textit{Low-energy effective theory.}---
Existing approaches based on the analysis of states at high-symmetry $k$-points in the BZ of a crystal are ruled out because momentum is no longer a good quantum number for quasicrystals. Instead, in the following we elucidate the HOTI in quasicrystals by establishing a low-energy effective theory in a continuum model and performing analysis in terms of coupled edge modes.

Generally, with the quasiperiodicity, the Fourier transform of any function $f(\mathbf{r})$ [e.g., the particle density $\rho(\mathbf{r})$ and quasicrystalline potential $U(\mathbf{r})$] of a 2D pentagonal quasicrystal can be expressed as
\begin{equation}
f(\mathbf{r})=\sum_{\mathbf{G} \in \mathcal{L}}\hat{f}(\mathbf{G})e^{i\mathbf{G}\cdot \mathbf{r}},
\end{equation}
where $\mathcal{L}$ is a countable set of reciprocal wave vectors that consist of  $\mathbf{G}=\sum_{j=1}^5 n_j \mathbf{g}_j$  ($n_j \in \mathbb{Z}$) filling densely the 2D reciprocal space. The five principal reciprocal vectors in the $k_x$-$k_y$ plane are
\begin{equation}
\mathbf{g}_j=2\pi(\cos[2\pi (j-1)/5],\sin[2\pi (j-1)/5]), \quad j=1, \ldots, 5,
\end{equation}
among which only four are independent due to the linear dependence of $\sum_{j=1}^5\mathbf{g}_j=0$. {Note that all the momentum values are scaled in the unit of inversed bond length $a^{-1}$, and we set $a=1$ without loss of generality.} The wave vectors in $\mathcal{L}$ can be divided into $n$-th order with $n=\sum_{j=1}^5|n_j|$, corresponding to the order of diffraction peaks [See Fig.~\ref{fig2}(a)]. The first order of $\mathcal{L}$, which contains ten principal wave vectors as $\pm \mathbf{g}_j$, defines the PBZ \cite{janssen2007aperiodic}, in analogy to the conventional first BZ of crystals. According to the gap labelling \cite{bellissard1992gap,Gambaudo_2014}, the leading-order gap opens at the boundary of the PBZ with the gap size determined by the first-order Fourier coefficient $\hat{U}(\pm \mathbf{g}_j)$ of quasicrystalline potential. Whereas nearby the $\Gamma$ point of the PBZ, the higher-order gaps appear in a hierarchy, which corresponds to multiple scattering processes having a rapidly decreasing size. This enables a continuum description based on a low-energy effective model with a proper truncation of $\mathcal{L}$.

Specifically, we employ the projection method, for which a quasicrystal can be viewed as a higher-dimensional crystal in hyperspace \cite{steurer126crystallography}. Then, an equivalent 4D representation of the 2D quasicrystal can be made by implementing the novel Fourier expansion proposed by Jiang and Zhang (JZ) \cite{JIANG2014428},
\begin{equation}
f(\mathbf{r})=\sum_{\mathbf{\Pi}} \hat{f}(\mathbf{\Pi}) e^{i\left[(\mathcal{S} \cdot \mathbf{\Pi})^{T} \cdot \mathbf{r}\right]},
\label{eq_ft}
\end{equation}
where $\mathbf{\Pi}= \sum_{i=1}^4 m_i \mathbf{Q}_i \in \mathbb{R}^{4}$, $m_i \in \mathbb{Z}$, and $\mathbf{Q}_i$ are the 4D primitive reciprocal vectors of the hypercrystal. $\mathcal{S}$ is the projection matrix connecting the 2D physical space with the 4D hyperspace. Mathematically, for any 2D quasiperiodic function $f(\mathbf{r})$, Eq.~(\ref{eq_ft}) has the property \cite{JIANG2014428}
\begin{equation}
\lim _{V \rightarrow \infty} \frac{1}{V} \int d \mathbf{r} f(\mathbf{r})=\left.\hat{f}(\mathbf{\Pi})\right|_{\mathbf{\Pi}=0}.
\end{equation}
Consequently, the large-scale average property of quasicrystals is well captured by the contribution around $\mathbf{k}=\mathcal{S} \cdot\mathbf{\Pi}=0$, i.e., the $\Gamma$ point of the PBZ. In fact, it has been proved that the behavior of a quasicrystal for excitations of any kind with long-wavelength modes can be related to its average structure \cite{PhysRevB.15.643,PhysRevB.34.596,PhysRevB.34.617, doi:10.1080/01418619808223760,PhysRevB.58.23,Wolny:js0074,Steurer:js0068,Aragonjs0118}.
Since the band inversion happens around $\Gamma$ (see Fig.~S1 in Supplemental Material \footnote{\label{fn}See Supplemental Material at http://link.aps.org/supplemental/xxx, for more details about the derivation of the effective Hamiltonian of the quasicrystal and numerical results, which include Refs.~\cite{JIANG2014428,steurer126crystallography,hiller1985crystallographic,RevModPhys.63.699,RevModPhys.64.3,PhysRevLett.76.1489,PhysRev.94.1498, PhysRevB.15.643,PhysRevB.34.596,PhysRevB.34.617,doi:10.1080/01418619808223760,PhysRevB.58.23,Wolny:js0074,Steurer:js0068,Aragonjs0118, doi:10.1143/JPSJ.77.031007,RevModPhys.83.1057,PhysRevLett.121.116801,PhysRevLett.98.186809,PhysRevB.82.115120,PhysRevB.99.155102,huanghqQLprl,huanghqQLprb}}),
a low-energy effective model at the long-wavelength limit, which can be approximated by the average structure of quasicrystals, is sufficient to describe the relevant topological physics.

To derive the effective Hamiltonian in the pseudo $k$-space, we first apply the JZ Fourier expansion to Eq.~(\ref{eq1}) without the Zeeman field, which yields (see Supplemental Material \footnotemark[\value{footnote}]),
\begin{eqnarray}
H(\mathbf{\Pi}) &=&\sum_{\alpha}\epsilon_{\alpha} c_{\mathbf{\Pi}, \alpha}^{\dagger} c_{\mathbf{\Pi}, \alpha}^{}+\sum_{ \alpha, \beta} t_{\alpha, \beta}(\mathbf{\Pi}) c_{\mathbf{\Pi}, \alpha}^{\dagger} c_{\mathbf{\Pi}, \beta}^{}\nonumber\\
&+&i \lambda \left(c_{\mathbf{\Pi}, p_{y}}^{\dagger} s_{z} c_{\mathbf{\Pi}, p_{x}}^{}-c_{\mathbf{\Pi}, p_{x}}^{\dagger} s_{z} c_{\mathbf{\Pi}, p_{y}}^{}\right).
\label{Hpi}
\end{eqnarray}
Here we have adopted the long-wavelength approximation to calculate an average hopping as
\begin{equation}
t_{\alpha,\beta}\left(\mathbf{\Pi}\right)\approx \lim _{V \rightarrow \infty} \frac{1}{V} \int d \mathbf{r} \mathcal{P}(\mathbf{r}) t_{\alpha, \beta}\left(\mathbf{r}\right) e^{i\left[(\mathcal{S} \cdot \mathbf{\Pi})^{T} \cdot \mathbf{r}\right]}.
\end{equation}
Note that around $\Gamma$, electron scattering is not affected by the local details of the quasicrystalline potential but only feels an average effective potential \cite{Aragonjs0118,Wolny:js0124}. Here $\mathcal{P}(\mathbf{r})$ is the statistical average distribution of interatomic vectors in quasicrystals, which is also known as the Patterson function and can be extracted from diffraction data \cite{PhysRev.46.372,PhysRevB.39.5850}. We then downfold the Hamiltonian to the two-orbital subspace around the Fermi level based on the L\"{o}wding perturbation method \cite{winkler2003soc,huang2013existence}, followed by projecting the two-orbital Hamiltonian to 2D space by taking $\mathbf{k}=\mathcal{S}\cdot\mathbf{\Pi}$ and expanding $\mathbf{k}$ around $\Gamma$. Finally, we obtain the effective $k \cdot p$ Hamiltonian without the Zeeman term as \footnotemark[\value{footnote}],
\begin{eqnarray}
H_{\mathrm{eff}}&=&\left(m-b k^{2}\right) \sigma_{z}+[a k_{y}+ a'(k_y^3+k_x^2k_y)]\sigma_{x} s_{z}\nonumber\\
 &+&[a k_{x}+a'(k_x^3+k_xk_y^2)] \sigma_{y},
\label{Heff}
\end{eqnarray}
where $\sigma$ and $s$ are Pauli matrices acting on the orbital and spin degrees of freedom, respectively. It is noteworthy that Eq.~(\ref{Heff}) still satisfies the $C_{5v}$ symmetry of the pentagonal quasicrystal. More importantly, its solution of band structure shows a band inversion between \textit{s} and \textit{p} orbitals around $\Gamma$ [see Fig.~\ref{fig2}(b)], indicating a nontrivial topology.

Now, let us consider a generic open boundary along {$\vec{e}_{1}=\cos \zeta \hat{e}_{x}+\sin \zeta \hat{e}_{y}$} direction with the normal vector {$\vec{e}_{2}=-\sin \zeta \hat{e}_{x}+\cos \zeta \hat{e}_{y}$} [see Fig.~\ref{fig2}(c)]. To further derive the low-energy Hamiltonian for the edge states along $\vec{e}_{1}$, we perform a rotational transformation to Eq.~(\ref{Heff}), and replace $k_{2} \rightarrow-i \partial_{x_{2}}, k_{1} \rightarrow 0$. After some algebra (see Supplemental Material \footnotemark[\value{footnote}]), we obtain a pair of edge-state solutions for two spin channels, and arrive at the 1D edge model to the leading order in $k_1$: $\mathcal{H}_{\mathrm{edge}}=-a k_{1} s_{z}$. It indicates the existence of a pair of spin-polarized gapless edge states protected by time-reversal symmetry.

When an in-plane Zeeman field of $H_{\mathrm{in}}^\prime= \delta_M(\cos\theta\sigma_0s_x+ \sin\theta \sigma_0s_y)$ is applied, the edge state $\mathcal{H}_{\mathrm{edge}}$ will generally be gapped by a mass term
\begin{equation}
\mathcal{M}(\zeta_i) \sim \cos \phi_i s_x+\sin \phi_i s_y,
\label{mass}
\end{equation}
where $\phi_i=\zeta_i +\theta -\pi /2 $ is the generalized phase of the effective Dirac mass \cite{PhysRevLett.98.186809,PhysRevB.82.115120}, which depends on the orientation angle $\zeta_i$ of the \textit{i}-th edge. From the work of Jackiw and Rebbi \cite{jackiw1976solitons}, a phase shift of $\Delta \phi=\Delta\zeta=\pi$ between two edges results in a CM with fractional charge $Q=e/2$ due to mass inversion. Moreover, according to Moore's theory~\cite{PhysRevB.99.155102}, a kink arising from the effective mass term at the corner gives rise to a localized CM with fractional charge of $Q=e|\Delta \phi / 2 \pi|=e|\Delta \zeta / 2 \pi|$ \cite{PhysRevLett.47.986,PhysRevLett.50.439}. Remarkably, for adjacent edges of a pentagonal Penrose-tiling quasicrystal, the angle difference $\Delta\zeta=2\pi/5$ leads to a fractional charge $Q=e/5$. On the contrary, an out-of-plane Zeeman field $H_{\mathrm{out}}^\prime= \delta_M\sigma_0s_z$ only contributes an energy-shift term of $\delta_M s_z$ which cannot gap the edge state $\mathcal{H}_{\mathrm{edge}}$ (see Supplemental Material \footnotemark[\value{footnote}]).

For the Ammann-Beenker-tiling quasicrystal, a similar low-energy theory can be derived and the phase shift becomes $\Delta\phi=\pi/4$ at corners of the octagonal sample, giving rise to a fractional charge of $Q=e/8$ at each corner. Thus, our model is valid for quasicrystals with odd- as well as even-rotational symmetries.
We stress that our approach of realizing CM in quasicrystals via the fractional mass kink is fundamentally different from previous works based on the mass-inversion mechanism \cite{PhysRevLett.124.036803,PhysRevLett.123.196401,PhysRevB.102.241102}, because their CMs rely on an alternating sign of the artificial mass term at the boundary, and hence does not work for odd rotations.

\begin{figure}
\includegraphics[width =1\columnwidth]{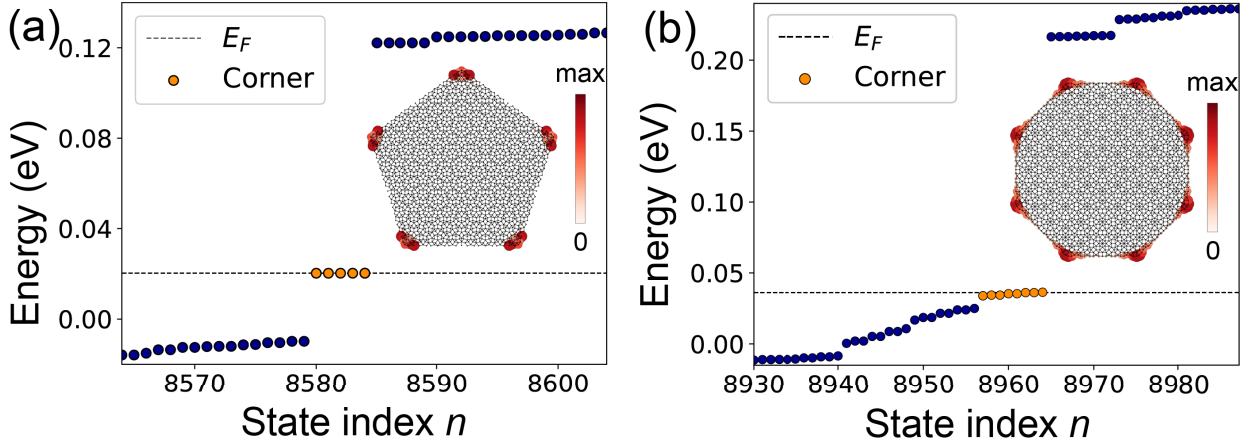}
\caption{\label{fig3}
Energy spectrum of a finite (a) pentagonal Penrose-tiling quasicrystal sample with 2146 atoms and (b) octagonal Ammann-Beenker-tiling quasicrystal sample with 2241 atoms in the presence of an in-plane Zeeman field with $\delta_M=0.12$ eV and $\delta_M=0.15$ eV, respectively. The parameters used here are $\epsilon_s=0.7, \epsilon_p=-2.3, V_{ss\sigma}=-V_{sp\sigma}=-0.17, 
V_{pp\sigma}=V_{pp\pi}=0.34$, 
and $\lambda=1.0$ eV. {Insets show the spatial distributions of CMs.}}
\end{figure}

\textit{Topological CMs in quasicrystals.}---
The above effective model predictions are confirmed by numerical TB calculations. In the absence of the Zeeman field, the model (\ref{eq1}) describes a TI state in the pentagonal quasicrystal, which is verified by the calculation results of a nonzero spin Bott index ($B_s=1$) \cite{huanghqQLprl,huanghqQLprb} and time-reversal symmetry-protected edge states residing inside the bulk gap (see Fig.~S3 and S4 in Supplemental Material \footnotemark[\value{footnote}]). In the presence of an in-plane Zeeman field along the \textit{x}-axis with $\delta_M=0.12$ eV, the energy spectrum of the finite pentagonal quasicrystal is gapped and five states appear at the Fermi level and separate from other states, as shown in Fig.~\ref{fig3}(a). We plot the spatial distribution of these states [see inset of Fig.~\ref{fig3}(a)], and found that they are indeed CMs localized at corners of the pentagonal quasicrystal. This implies that the system becomes a HOTI, although the CMs are not located at the mid-gap position due to the lack of chiral or particle-hole symmetry. Remarkably, since four of the five CMs are occupied for a charge-neutral system, a fractional charge of $e/5$ localized at each corner can be realized by adding one electron, resulting in a fractionalized charge distribution due to the filling anomaly \cite{PhysRevB.99.245151,physrevb.103.205123}. Similarly, for the octagonal Ammann-Beenker-tiling quasicrystal, we found eight CMs in the gap and a charge fractionalization of $e/8$ per corner if one extra electron is added [see Fig.~\ref{fig3} (b)].
It is worth noting that the corner charge in crystals are fractionally quantized module of $e/n$ with $n=2,3,4$ and $6$, covering all the allowed rotational symmetries by the crystallographic restriction theorem. Our results extend the possible values of the corner charge fractionalization with $e/q$, where $q=5,8$ and other rotational orders in quasicrystals.

We further investigated numerically the phase evolution with the field strength $\delta_M$ {for the Penrose-tiling quasicrystal}. By increasing $\delta_M$, the bulk energy gap, which is estimated from the calculations of quasicrystal approximants with 1364 atoms \footnotemark[\value{footnote}], decreases gradually and closes eventually when $\delta_M>0.2$ eV. Whilst the CMs exist in the gap only when $\delta_M<0.15$ eV, beyond which they merge into the bulk spectrum. In addition, we studied the effect of arbitrary Zeeman field orientations in the \textit{x}-\textit{y} plane, e.g., $\bm{m}=(\cos \theta,\sin \theta, 0)$, and found that the CMs persist regardless of $\theta$. This can be understood by simply performing a rotation of spin about the \textit{z} axis to make the Zeeman field pointing along the $x$ direction \cite{PhysRevLett.124.166804}.

\begin{figure}
\includegraphics[width =1.\columnwidth]{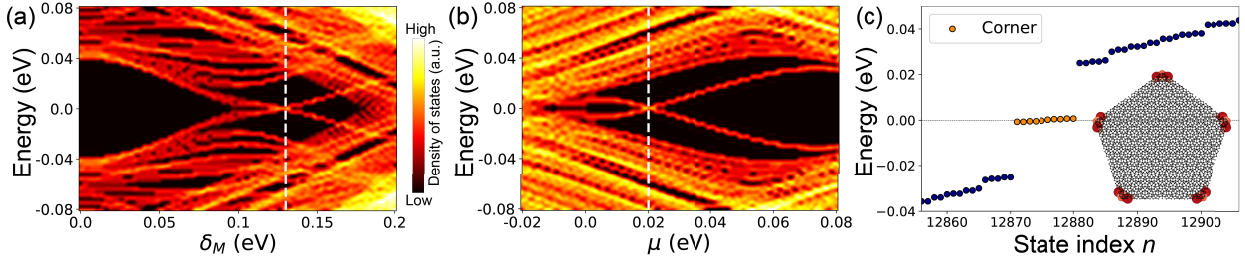}%
\caption{\label{fig4} Majorana CMs in the quasicrystal TI/\textit{s}-wave SC heterostructure with an in-plane Zeeman field.
(a) Quasiparticle energy spectrum vs the Zeeman-field strength $\delta_M$ at fixed $\Delta_0=0.04$ eV and $\mu=0.02$ eV. (b)
Same as (a) vs the chemical potential $\mu$ at fixed $\Delta_0=0.04$ eV and $\delta_M=0.13$ eV. (c) Energy spectrum of a finite pentagonal sample at $(\delta_M,\mu,\Delta_0)=(0.13,0.02,0.04)$ eV. {Other parameters are the same as Fig.~\ref{fig3}.} The inset shows the real-space distribution of Majorana CMs.}
\end{figure}

\textit{Majorana CMs in quasicrystal TI and \textit{s}-wave SC heterostructures.}---
In addition to fractional topological corner charge, our model and approach are also applicable to create Majorana CMs in the TI quasicrystals, under an in-plane Zeeman field and in proximity with \textit{s}-wave SCs. The physics of the heterostructure can be described by an effective Bogoliubov-de Gennes (BdG) Hamiltonian
\begin{equation}
H_{\mathrm{BdG}}=\left(
          \begin{array}{cc}
            H-\mu & \Delta_0 \\
            \Delta_0^{\dagger} & -H^*+\mu \\
          \end{array}
        \right),
\end{equation}
where $\mu$ is the chemical potential and $\Delta_0$ denotes \textit{s}-wave SC pairing gap by proximity effect. Although the normal-state part of $H_{\mathrm{BdG}}$ is topologically nontrivial with helical edge states, the proximity induced \textit{s}-wave SC pairing necessarily trivializes the full BdG model and gaps out the edge states \cite{PhysRevLett.123.167001}. As shown in Fig.~\ref{fig4}(a), with the increasing in-plane Zeeman field $\delta_M$ at a fixed $\Delta_0$, the quasiparticle energy gap first closes and then reopens accompanied by the emergence of localized modes at corners, implying the nontrivial topology is resumed. Moreover, these CMs can be fine-tuned to zero energy by adjusting the chemical potential, giving rise to Majorana CMs \cite{PhysRevB.100.205406} [see Fig.~\ref{fig4}(b)]. As shown in Fig.~\ref{fig4}(c), in the topological phase at $(\delta_M,\mu,\Delta_0)=(0.13,0.02,0.04)$ eV, five pairs of Majorana zero modes emerge inside the gap. The inset of Fig.~\ref{fig4}(c) shows the spatial distribution of these zero modes, confirming that they are localized around the corners of a finite pentagonal quasicrystal.

\textit{Conclusion.}---
We have devised a low-energy theory of quasicrystals under the long-wavelength approximation, and demonstrated that higher-order topological CMs with fractional charges can be generated by in-plane Zeeman-field-induced fractional mass kinks. Our model predictions are further confirmed by numerical TB calculations, which show also emergence of Majorana CMs in TI quasicrystals in proximity with an \textit{s}-wave SC. Our work greatly extends the higher-order topological physics for mass-kink induced domain-wall states to quasicrystals and establishes a generic theoretical framework to study the low-energy physics of quasicrystalline systems.

\begin{acknowledgments}
This work was supported by the National Key R\&D Program of China (No. 2021YFA1401600), the National Natural Science Foundation of China (Grant No. 12074006), and the start-up fund from Peking University. F.L. acknowledges the support of U.S. DOE-BES (Grant No. DE-FG02-04ER46148). The computational resources were supported by the high-performance computing platform of Peking University.
\end{acknowledgments}

\providecommand{\noopsort}[1]{}\providecommand{\singleletter}[1]{#1}%

\end{document}